# On the Relativistic Origin of Inertia and Zero-Point Forces

Charles T. Ridgely
charles@ridgely.ws

Current approaches to the problem of inertia attempt to explain the inertial properties of matter by expressing the inertial mass appearing in Newton's second law of motion in terms of some other more fundamental interaction. One increasingly popular approach explains inertial and gravitational forces as drag forces arising due to quantum vacuum zero-point phenomena. General relativity, however, suggests that gravitational and inertial forces are manifestations of space-time geometry. Based on this, the present analysis demonstrates that inertia and zero-point forces are ultimately relativistic in origin. Additionally, it is argued that body forces induced on matter through zero-point interactions are resistive forces acting in addition to inertia.

## 1. Introduction

One of the longest standing questions in physics certainly has been by what means do material bodies resist changes to their states of motion (inertia). One approach to this problem has been to view inertia as merely a fundamental property of all matter with no further explanation attainable. Another approach has been to view inertia as arising due to a gravitational coupling among all matter in the universe. This approach, first proposed by Mach (ca. 1883), stems from the notion that relative motion is meaningless in the absence of surrounding matter. With this in mind, one is then led to the idea that the inertial properties of matter must somehow be related to the cosmic distribution of all matter in the universe [1]. Unfortunately, Mach's principle leads to action at a distance phenomena which even today appear irreconcilable with accepted theory.

More recent studies have sought to explain inertia by expressing the inertial mass $m$ appearing in Newton's second law of motion,

$$\mathbf{f} = \frac{d}{dt}(m\mathbf{v}) \qquad (1)$$

in terms of some other more fundamental entity or interaction. At present, there appears to be several approaches to this end. One approach put forth by J. F. Woodward and T. Mahood seeks to preserve, and apparently expand upon, the Machian view mentioned above [1]-[2]. Another approach, put forth within the Standard Model of particle physics, proposes a scalar Higgs field which assigns specific quantities of energy to elementary particles. With this approach, elementary particles possess inertial properties simply because they possess energy, which is equivalent to mass. Unfortunately, the Higgs mechanism stops short of actually explaining why mass, or energy, resists acceleration. Consequently, the Higgs mechanism leaves us with the above-mentioned notion that inertia is a fundamental property of all matter with no further explanation attainable. Still another approach to the problem of inertia has been put forth by B. Haisch, A. Rueda, H. E. Puthoff, et al [3]-[4]. With their approach, it is proposed that the inertial properties of ordinary matter are due to local interactions between a vacuum electromagnetic zero-point field (ZPF) and subatomic particles, such as quarks and electrons, constituting ordinary matter. In essence, the ZPF approach asserts that when a material object accelerates through the zero-point radiation pervading all of space, some of the radiation is scattered by the quarks and electrons constituting the object. This scattering of radiation exerts a reactive body force on the object, which, according to the ZPF proposal, can be associated with the inertial mass of the object.

Although we challenged the ZPF proposal in previous papers [5]-[6], the ZPF proposal seems increasingly appealing. Not only does the ZPF proposal suggest a local basis for gravitational and inertial forces, thus avoiding action at a distance phenomena often associated with Mach's principle, but also seems to suggest an intimate relationship between electromagnetism, inertia, and gravitation [3]-[4]. However, while current ZPF theory does show that zero-point phenomena give rise to body forces on ordinary matter, this is only part of the story.

According to the ZPF proposal, the intrinsic rest mass-energy content of matter is induced in large part through interaction with zero-point radiation. Incident zero-point radiation imparts an ultrarelativistic jittering motion, or *zitterbewegung* [4], to the quarks and electrons comprising ordinary matter, thereby endowing matter with large quantities of internal kinetic energy. With this paradigm, the rest mass-energy of matter predicted by the familiar expression $E = mc^2$ is interpreted as entirely ZPF-induced kinetic energy. But this seems to suggest that subatomic particles have no intrinsic rest mass-energy aside from their ZPF-induced motion. Of course subatomic particles are, indeed, well known to possess definite quantities of intrinsic rest mass-energy. A more traditional interpretation is that $E = mc^2$ is simply a statement that all forms of energy exhibit inertial properties [7]-[8], which can be associated with inertial mass, and that mass is merely one particular embodiment of energy. Herein, both interpretations are utilized; the energy content of matter comprises intrinsic rest mass-energy as well as internal kinetic energy due to ZPF-induced zitterbewegung [4] of subatomic particles.

Another important point to recognize is that general relativity suggests that gravitational and inertial forces alike arise due to the behavior of space-time [9]-[14]. This implies that greater insight into the origin of inertia cannot be obtained simply by eliminating the inertial mass $m$ appearing in Eq. (1) in favor of some other entity or interaction. Rather, space-time is a fundamental participant in the generation of inertial and gravitational forces, and as such must be taken into account, as well [5]. Thus, while the energy content of matter may indeed be largely ZPF-induced, one must still explain why energy resists acceleration.

## 2. Proper Force Experienced by a Uniformly Accelerating Observer

Let a first, moving observer be accelerating uniformly under the action of a constant, external force applied by a second observer residing within a stationary system $S$. From the vantage of the accelerating reference system, $S'$, the moving observer's local coordinate system appears as though characterized by the presence of a uniform gravitational field [15]. The line element in the local accelerating coordinate system can be expressed generally in the form

$$ds^2 = c^2 g_{00}(x,y,z) dt^2 - dx^2 - dy^2 - dz^2 \quad (2)$$

in which $g_{00}(x,y,z)$ contains information about the force on the moving observer. So long as the moving observer's acceleration is uniform, $g_{00}(x,y,z)$ remains independent of time [16]-[17]. Provided that the moving observer remains stationary at the origin of the accelerating coordinate system at all times, Eq. (2) leads to

$$g_{00}(x,y,z) = \left(\frac{d\tau}{dt}\right)^2 \quad (3)$$

In this expression, $d\tau$ is an interval of proper time experienced by the accelerating observer and $dt$ is an interval of time experienced by a momentarily comoving observer whose coordinate origin is momentarily coincident with that of the accelerating observer at a time $t = 0$.

The proper force experienced by the moving observer while accelerating can be derived by using the geodesic equation:

$$\frac{d^2 x^\alpha}{d\tau^2} + \Gamma^\alpha_{\mu\nu} \frac{dx^\mu}{d\tau} \frac{dx^\nu}{d\tau} = 0 \quad (4)$$

in which the Christoffel symbol $\Gamma^\alpha_{\mu\nu}$ is given by

$$\Gamma^\alpha_{\mu\nu} = \frac{1}{2} g^{\alpha\beta} \left(g_{\mu\beta,\nu} + g_{\beta\nu,\mu} - g_{\mu\nu,\beta}\right) \quad (5)$$

Using Eq. (2) for the case of the moving observer remaining stationary at the origin of the accelerating coordinate system, Eq. (4) reduces to

$$\frac{d^2 x^\alpha}{d\tau^2} + c^2 \Gamma^\alpha_{00} \left(\frac{dt}{d\tau}\right)^2 = 0 \quad (6)$$

Expressing the Christoffel symbol in terms of the accelerating coordinate system then leads to

$$\frac{d^2 x^\alpha}{d\tau^2} - \frac{c^2}{2} \left(\frac{dt}{d\tau}\right)^2 g^{\alpha j} g_{00,j} = 0 \quad (7)$$

wherein Latin indices are carried over the set of values {1, 2, 3}. This expression simplifies further upon noticing that partial differentiation of Eq. (3) yields

$$g_{00,j} = 2\left(\frac{d\tau}{dt}\right) \partial_j \left(\frac{d\tau}{dt}\right) \quad (8)$$

in which $\partial_j$ denotes partial differentiation with respect to $x^j$. As a further simplification, the inverse metric tensor $g^{\alpha j}$ can be expressed in the form

$$g^{\alpha j} = \begin{cases} 0, & \alpha = 0 \\ -\delta^{ij}, & \alpha = i \end{cases} \quad (9)$$

wherein $\delta^{ij}$ equals unity when $i = j$, and equals zero when $i \neq j$. Substituting Eqs. (8) and (9) into Eq. (7) leads to

$$\frac{d^2 x^j}{d\tau^2} + c^2 \delta^{ij} \frac{dt}{d\tau} \partial_j \left(\frac{dt}{d\tau}\right) = 0 \quad (10)$$

Next, expressing the second term in terms of a natural logarithm, Eq. (10) can be recast in the form

$$\frac{d^2 x^j}{d\tau^2} - c^2 \delta^{ij} \partial_j \left\{\ln\left(\frac{dt}{d\tau}\right)\right\} = 0 \quad (11)$$

This is the proper acceleration experienced by the moving observer.

Equation (11) can be used to derive an expression for the force experienced by the moving observer upon noticing that the external force acting on the observer is expressible as

$$f^i = m \frac{d^2 x^i}{d\tau^2} \quad (12)$$

where $m$ is the moving observer's proper mass. Using Eq. (12), and simplifying a bit, Eq. (11) then becomes

$$f^i = mc^2 \delta^{ij} \partial_j \left\{\ln\left(\frac{dt}{d\tau}\right)\right\} \quad (13)$$

This is the proper force experienced by the moving observer while accelerating under the external force applied by the stationary observer in $S$ [5].

From Eq. (13), the inertial resistance of the accelerating observer experienced by the stationary observer can be derived. One approach is to transform Eq. (13) from the accelerating system $S'$ to the stationary system $S$. Another approach is to consider the case of weak acceleration, thereby placing the proper force within the Newtonian realm, and then appeal to Newton's third law of motion. Choosing the latter approach, one will notice that when the moving observer accelerates weakly, the scalar function $dt/d\tau$ assumes values very close to unity. For the case of weak acceleration, therefore, the natural logarithm of the scalar function $dt/d\tau$ may be approximated by use of the expression [18]

$$\ln\left(\frac{dt}{d\tau}\right) \approx \frac{dt}{d\tau} - 1 \qquad (14)$$

Using this approximation in Eq. (13) leads directly to

$$f^i = mc^2 \delta^{ij} \partial_j \left(\frac{dt}{d\tau}\right) \qquad (15)$$

This is the force experienced by the moving observer in the limit of weak, Newtonian acceleration. Expressing Eq. (15) in vector notation, the force experienced by the moving observer becomes

$$\mathbf{f} = E\nabla\left(\frac{dt}{d\tau}\right) \qquad (16)$$

where $E = mc^2$ has been used, and $\nabla$ represents the three dimensional gradient operator. Equation (16) expresses the force imparted to the moving observer by the stationary observer within system $S$. Assuming that the stationary system $S$ is an inertial frame, Newton's third law of motion suggests that the reaction force experienced by the stationary observer is given by $\mathbf{f}_{in} = -\mathbf{f}$. The subscript on $\mathbf{f}_{in}$ indicates that the reaction force is due to the inertia of the accelerating observer. Using this force with Eq. (16) leads directly to

$$\mathbf{f}_{in} = -E\nabla\left(\frac{dt}{d\tau}\right) \qquad (17)$$

This is the inertial resistance of the accelerating observer experienced by the stationary observer in $S$ [5]. It is straightforward to see that the inertial resistance is solely dependent upon the total proper energy content of the accelerating observer and the gradient of the scalar function $dt/d\tau$ arising due to relativity. The form of Eq. (17) suggests that all forms of energy resist acceleration and possess inertial properties which are entirely relativistic in origin [5],[7]-[8]. On this basis, it may be concluded that inertia is purely relativistic.

## 3. Inertial Resistance of an Observer Accelerating in Flat Space-Time

In the previous Section, general relativity was used to derive the proper force experienced by a uniformly accelerating observer. In this Section, a purely special relativistic approach is used to derive the inertial resistance of an observer accelerating uniformly through flat space-time.

Consider a stationary observer residing in an inertial system $S$ who applies a force to a second, moving observer such that the moving observer accelerates tangentially along a circular path of radius $r$. It is envisioned that the stationary observer exerts the force on the moving observer through the use of some mechanical means, which exerts a constant torque on the moving observer. The stationary observer compares the lengths of time intervals in the moving and stationary systems each time the moving observer reaches the point of closest approach; that is, when the relative velocity is entirely transverse. As the moving observer's velocity increases, the stationary observer finds that time in the moving system becomes increasingly dilated according to the familiar expression [19]:

$$\left(\frac{dt}{d\tau}\right) = \frac{1}{\sqrt{1 - v^2/c^2}} \qquad (18)$$

where $dt$ is a time interval in system $S$ and $d\tau$ is a corresponding interval of proper time experienced by the moving observer. The stationary observer determines an initial time dilation $(dt/d\tau)_i$; and then at some later time, the stationary observer determines a final time dilation $(dt/d\tau)_f$:

$$\left(\frac{dt}{d\tau}\right)_i = \frac{1}{\sqrt{1 - v_i^2/c^2}}, \quad \left(\frac{dt}{d\tau}\right)_f = \frac{1}{\sqrt{1 - v_f^2/c^2}} \qquad (19a, b)$$

where $v_i$ and $v_f$ are initial and final velocities of the moving observer at the instants when the initial and final time dilation are respectively measured.

The inertial resistance of the moving observer can be derived by first using Eqs. (19) to express the change in time dilation as

$$\left(\frac{dt}{d\tau}\right)_f - \left(\frac{dt}{d\tau}\right)_i = \frac{1}{\sqrt{1 - v_f^2/c^2}} - \frac{1}{\sqrt{1 - v_i^2/c^2}} \qquad (20)$$

in which the magnitude of the moving observer's tangential acceleration remains constant. In order to derive the inertial resistance by use of Eq. (20), one will notice that the change in the scalar function $dt/d\tau$ on the left hand side of Eq. (20) may be expressed as a line integral of the form

$$\left(\frac{dt}{d\tau}\right)_f - \left(\frac{dt}{d\tau}\right)_i = \int_i^f \nabla\left(\frac{dt}{d\tau}\right) \cdot d\mathbf{l} \qquad (21)$$

where $d\mathbf{l}$ is an infinitesimal length vector aligned along the circular path on which the moving observer travels. Also, the right-hand side (RHS) of Eq. (20) can be expressed as

$$\frac{1}{\sqrt{1 - v_f^2/c^2}} - \frac{1}{\sqrt{1 - v_i^2/c^2}} = \frac{1}{c^2}\int_i^f \frac{\mathbf{v} \cdot d\mathbf{v}}{\left(1 - v^2/c^2\right)^{3/2}} \qquad (22)$$

where $v$ is an instantaneous velocity of the moving observer at a point between the velocities $v_i$ and $v_f$. Using Eqs. (21) and (22) in Eq. (20) gives

$$\int_i^f \nabla\left(\frac{dt}{d\tau}\right) \cdot d\mathbf{l} = \frac{1}{c^2}\int_i^f \frac{\mathbf{v} \cdot d\mathbf{v}}{\left(1 - v^2/c^2\right)^{3/2}} \qquad (23)$$

Upon expanding the RHS of this expression by use of the relation

$$\frac{1}{\left(1 - v^2/c^2\right)^{3/2}} = \frac{1}{\sqrt{1 - v^2/c^2}} + \frac{(\mathbf{v} \cdot \mathbf{v})}{c^2\left(1 - v^2/c^2\right)^{3/2}} \qquad (24)$$

and performing some algebraic manipulation, Eq. (23) then becomes

$$\int_i^f \nabla\left(\frac{dt}{d\tau}\right) \cdot d\mathbf{l} = \frac{1}{c^2}\int_i^f \mathbf{v} \cdot \left[\frac{d\mathbf{v}}{\sqrt{1-v^2/c^2}} + \frac{\mathbf{v}(\mathbf{v}\cdot d\mathbf{v})}{c^2\left(1-v^2/c^2\right)^{3/2}}\right] \quad (25)$$

This expression can be simplified since

$$\frac{d\mathbf{v}}{\sqrt{1-v^2/c^2}} + \frac{\mathbf{v}(\mathbf{v}\cdot d\mathbf{v})}{c^2\left(1-v^2/c^2\right)^{3/2}} = d\left(\frac{\mathbf{v}}{\sqrt{1-v^2/c^2}}\right) \quad (26)$$

Using this, and expressing the velocity of the moving observer as $\mathbf{v} = d\mathbf{l}/dt$, Eq. (25) can be rewritten in the form

$$\int_i^f \nabla\left(\frac{dt}{d\tau}\right) \cdot d\mathbf{l} = \frac{1}{mc^2}\int_i^f \frac{d\mathbf{l}}{dt} \cdot d\left(\frac{m\mathbf{v}}{\sqrt{1-v^2/c^2}}\right) \quad (27)$$

in which the moving observer's proper mass, $m$, has been introduced on the RHS. Rearranging Eq. (27) leads to

$$\int_i^f \nabla\left(\frac{dt}{d\tau}\right) \cdot d\mathbf{l} = \frac{1}{mc^2}\int_i^f \frac{d}{dt}\left(\frac{m\mathbf{v}}{\sqrt{1-v^2/c^2}}\right) \cdot d\mathbf{l} \quad (28)$$

It is now straightforward to see that the integrand on the RHS of this expression contains the familiar relativistic generalization of Newton's second law of motion [20]-[21]. Taking this into account, Eq. (28) can be recast in the form

$$\int_i^f \nabla\left(\frac{dt}{d\tau}\right) \cdot d\mathbf{l} = \frac{1}{E}\int_i^f \mathbf{f} \cdot d\mathbf{l} \quad (29)$$

wherein $E = mc^2$ has been used, and $\mathbf{f}$ is the external force imparted to the moving observer by the stationary observer.

While the moving observer accelerates, there are at least two forces detectable by the stationary observer. One force is the external force $\mathbf{f}$ applied to the moving observer by the stationary observer. A second force is the moving observer's inertial resistance $\mathbf{f}_{in}$. Assuming that no other external forces are present, the inertial resistance is related to the external force according to $\mathbf{f}_{in} = -\mathbf{f}$. Substituting this into Eq. (29), and eliminating the integration on both sides of the expression leads directly to

$$\mathbf{f}_{in} = -E\nabla\left(\frac{dt}{d\tau}\right) \quad (30)$$

This is the inertial resistance of the moving observer, experienced by the stationary observer residing in flat space-time [5]. The form of Eq. (30) is identical to that of Eq. (17), derived on the basis of general relativity, and suggests that the inertial resistance of the moving observer arises due to the anisotropy of space-time within the accelerating system [5]. On this basis, we deduce that all forms of energy, regardless of embodiment, exhibit inertial properties arising as purely relativistic phenomena [7]-[8].

## 4. Resistance Force on Accelerating Matter due to Zero-Point Radiation

According to the preceding Sections, special and general relativity suggest that the inertial properties of matter, as well as of all other forms of energy, are purely relativistic in origin. In this Section, it is demonstrated that forces induced on matter through interaction with zero-point radiation are ultimately relativistic in origin, and that such forces arise in addition to inertial forces. It should be pointed out that the expression "zero-point radiation" as used herein is intended to reference all frequency modes comprising the zero-point radiation field (ZPF), and is not meant to focus in on any one particular mode thereof.

Consider a block of matter, of proper volume $V$, undergoing uniform acceleration through flat space-time due to an external force. Observers residing on the block detect a flux of zero-point radiation entering the block through the front side, which may be called wall $A$, and passing out of the block through the opposite side, called wall $B$. According to these observers, zero-point radiation within the volume of the block possesses a momentum density of the form

$$\Delta\mathbf{p} = \frac{1}{c^2}(\mathbf{S}_A - \mathbf{S}_B) \quad (31)$$

where $\mathbf{S}_A$ and $\mathbf{S}_B$ are Poynting vectors corresponding to the flux of zero-point radiation detected at walls $A$ and $B$, respectively. While the block accelerates, the Poynting vectors are not equal in magnitude; zero-point radiation becomes Doppler-shifted as it passes through the accelerating frame of the block. Taking this into consideration, the Poynting vectors may be expressed in terms of the energy density of the ZPF observed at each wall:

$$\mathbf{S}_A = cu_A\mathbf{n}, \quad \mathbf{S}_B = cu_B\mathbf{n} \quad (32a, b)$$

where the subscripts $A$ and $B$ indicate the walls at which the energy density of the ZPF is detected, and $\mathbf{n}$ is a unit vector pointing in the direction of the block's acceleration. Since the block is accelerating, zero-point radiation gains energy as it passes from wall $A$ to wall $B$. As a result, radiation detected at wall $B$ appears blue-shifted relative to radiation detected at wall $A$. Taking this into account, the Poynting vector given by Eq. (32b) can be expressed as

$$\mathbf{S}_B = cu_A\left(\frac{d\tau_A}{d\tau_B}\right)^2\mathbf{n} \quad (33)$$

where $d\tau_A$ and $d\tau_B$ are intervals of proper time experienced by the observers situated at walls $A$ and $B$, respectively. Using Eqs. (32a) and (33) in Eq. (31), the momentum density of zero-point radiation within the block becomes

$$\Delta\mathbf{p} = \frac{u_A}{c}\left[1 - \left(\frac{d\tau_A}{d\tau_B}\right)^2\right]\mathbf{n} \quad (34)$$

Expressing this in terms of time experienced by observers residing in a momentarily comoving reference frame (MCRF), and simplifying, leads to

$$\Delta \mathbf{p} = \frac{u_A}{c} \left(\frac{d\tau_A}{dt}\right)^2 \left[\left(\frac{dt}{d\tau_A}\right)^2 - \left(\frac{dt}{d\tau_B}\right)^2\right] \mathbf{n} \qquad (35)$$

where $dt$ is an interval of time experienced by the observers in the MCRF.

From Eq. (35), the force density due to zero-point radiation passing through the block can be derived by using $\mathbf{f} = \Delta \mathbf{p}/\Delta \tau$, where $\Delta \tau$ is an interval of proper time in the accelerating frame of the block. Supposing that the $x'$–axis of the accelerating coordinate system is oriented in the direction of the block's acceleration, observers positioned at wall $A$ can express $\Delta \tau$ in terms of the block's length, $\Delta x'$, along the $x'$–axis of the accelerating frame. This is carried out by observing that the time taken for a light signal to complete a round trip across the block is $\Delta \tau = 2\Delta x'/c$. Using this time interval and Eq. (35), the force density can then be expressed as

$$\mathbf{f} = \frac{u_A}{2\Delta x'} \left(\frac{d\tau_A}{dt}\right)^2 \left[\left(\frac{dt}{d\tau_A}\right)^2 - \left(\frac{dt}{d\tau_B}\right)^2\right] \hat{\mathbf{x}}' \qquad (36)$$

where $\hat{\mathbf{x}}'$ is a unit vector aligned along the $x'$–axis of the local accelerating coordinate system. Equation (36) holds for volumes in which $d\tau_A \neq d\tau_B$ and $\Delta x'$ assumes measurable values. When the volume of the block is on the order of a particle-sized volume, however, then $d\tau_A$ is approximately equal to $d\tau_B$ and $\Delta x'$ is infinitesimally small. The force density within a particle-sized volume can be obtained by taking the limit of Eq. (36) as $d\tau_B$ tends to $d\tau_A$ and $\Delta x'$ tends to zero:

$$\mathbf{f}' = \lim_{\substack{d\tau_B \to d\tau_A \\ \Delta x' \to 0}} \mathbf{f} \qquad (37)$$

Carrying out the limit for the case of a particle and noting that $dt/d\tau_B > dt/d\tau_A$, the force density assumes the form [5]-[6]

$$\mathbf{f}' = -u \frac{\partial}{\partial x'}\left[\ln\left(\frac{dt}{d\tau}\right)\right] \hat{\mathbf{x}}' \qquad (38)$$

where $u$ is the proper energy density of the ZPF according to observers moving with the particle.

Equation (38) expresses the force density due to zero-point radiation passing through and interacting with a particle of matter undergoing substantial acceleration. It will be noticed that Eq. (38) is substantially similar to Eq. (13) of Section 2. In Section 2, it was pointed out that when the acceleration is weak, the scalar function $dt/d\tau$ can be approximated by use of the expression given by Eq. (14). Upon using Eq. (14) with the force density, embodied by Eq. (38), one immediately obtains [5]-[6]

$$\mathbf{f}' = -u_0 \frac{\partial}{\partial x'}\left(\frac{dt}{d\tau}\right) \hat{\mathbf{x}}' \qquad (39)$$

Upon transforming this expression from the accelerating system to the inertial frame from which the external force on the particle arises, the force density becomes [5]-[6]

$$\mathbf{f} = -u\nabla\left(\frac{dt}{d\tau}\right) \qquad (40)$$

where the prime has been dropped from $\mathbf{f}'$ for simplicity. Equation (40) is a resistance force on the particle due to interaction with zero-point radiation.

According to a force-producing observer residing in flat space-time, when an external force is applied to the particle, a resistance force $\mathbf{f}_{ZPF}$ arises due to interaction between the particle and zero-point radiation. Using Eq. (40), such as observer can express this resistance force as

$$\mathbf{f}_{ZPF} = -uV\nabla\left(\frac{dt}{d\tau}\right) \qquad (41)$$

where $V$ is the proper volume of the accelerating particle. Equation (41) gives the resistance force acting on the accelerating particle due to the scattering of zero-point radiation. According to Eq. (41), the force on the particle arises not only due to the presence of zero-point radiation, but also due to the gradient of the scalar function $dt/d\tau$. It is interesting to note that when the scalar function $dt/d\tau$ assumes a constant value (i.e., the particle is no longer accelerating), the force given by Eq. (41) is then zero. This suggests that while zero-point phenomena may give rise to a large portion of the mass-energy content of matter, the inertia of such energy is purely relativistic [5].

At this point, it is worthwhile to show that Eq. (41) can be used to derive an expression for the total resistance force acting on the accelerating particle discussed above. According to the ZPF proposal, when a material body undergoes acceleration due to an external force, the quarks and electrons constituting the body scatter a portion of the zero-point radiation passing through the body. The scattered portion of radiation imparts an energy density to the accelerating body, which is expressed in the form [4]

$$u_{ZPF} = \int \eta(\omega) \frac{\hbar\omega^3}{2\pi^2 c^3} d\omega \qquad (42)$$

In this expression, $\eta(\omega)$ is a spectral function that governs the fraction of zero-point radiation that actually interacts with the accelerating body. Those who support the ZPF proposal interpret the energy density, given by Eq. (42), as the sole origin of the inertial mass of matter [3]-[4]. However, as discussed in the Introduction, another form of energy that must be taken into account is the intrinsic proper mass-energy of the accelerating body [5]-[8]. On this basis, the total energy content of the body must be expressed as $E_{Total} = E + E_{ZPF} + U$, where $E$ is the intrinsic mass-energy of the body, $E_{ZPF}$ is the internal kinetic energy due to ZPF-induced zitterbewegung [4] of the subatomic particles comprising the body, and $U$ includes additional forms of energy that may be present.

Using this expression for the accelerating particle discussed above leads to

$$E_{Total} = mc^2 + V\int \eta(\omega) \frac{\hbar\omega^3}{2\pi^2 c^3} d\omega + U \qquad (43)$$

where $E = mc^2$ has been used in the first term, Eq. (42) has been used to obtain the second term, and $V$ is the proper volume of the particle.

In order to derive an expression for the total resistance force acting of the accelerating particle, Eq. (41) must be amended to include all forms of energy possessed by the particle [7]-[8]. This calls for replacing $uV$ in Eq. (41) with the total energy given by Eq. (43). Carrying this out gives the force in the form

$$\mathbf{f} = -\left(mc^2 + V\int \eta(\omega)\frac{\hbar\omega^3}{2\pi^2 c^3}d\omega + U\right)\nabla\left(\frac{dt}{d\tau}\right) \qquad (44)$$

This expresses the total resistance force acting on the uniformly accelerating particle.

Equation (44) can be simplified by considering the case of weak, uniform acceleration along the $x$–coordinate axis of an inertial system. For this case, the scalar function $dt/d\tau$ may be expressed in the form

$$\frac{dt}{d\tau} \approx 1 + \frac{ax}{c^2} \qquad (45)$$

where $a$ is the acceleration and $x$ is a distance traveled along the $x$–axis due to the acceleration. Carrying out the gradient of this expression leads to

$$\nabla\left(\frac{dt}{d\tau}\right) = \frac{\mathbf{a}}{c^2} \qquad (46)$$

where the acceleration is expressed in vector notation as $\mathbf{a} = a\hat{\mathbf{x}}$. Substituting Eq. (46) into Eq. (44), and simplifying, then leads to

$$\mathbf{f} = -m\mathbf{a} - \left(\frac{V}{c^2}\int \eta(\omega)\frac{\hbar\omega^3}{2\pi^2 c^3}d\omega\right)\mathbf{a} - \frac{U}{c^2}\mathbf{a} \qquad (47)$$

Equation (47) is the total resistance force acting on a particle of proper mass $m$ that accelerates uniformly through zero-point radiation. With the exclusion of the first and last terms, Eq. (47) is identical to an expression for the force derived on the basis of ZPF theory [4].

## 5. Discussion and Conclusions

The present analysis has shown that the inertial properties of ordinary matter are ultimately relativistic in origin [5]. General relativity asserts that the forces of gravitation and inertia arise directly out the structure of space-time [9]-[14]. Based on this, both special and general relativity were used to derive an expression for the inertial resistance of an observer undergoing acceleration due to an external force. Both approaches led to an expression for the inertial resistance suggesting that all forms of energy possess inertial properties manifesting chiefly due to the anisotropy of space-time within accelerating systems [5]. Although electromagnetic zero-point phenomena may well be the origin of the energy content of matter [4], neither of the above-mentioned approaches required a specific appeal to zero-point phenomena. On this basis, it was concluded that inertia is a purely relativistic manifestation [5].

The present analysis demonstrated that forces induced on matter due to zero-point radiation are additional body forces arising due to the relativistic properties of space-time within accelerating systems. In addition, the expression for the force was used in conjunction with an expression for the energy density of zero-point radiation that interacts with an accelerating particle [4]. This led to an expression suggesting that ZPF-induced forces are resistive forces acting in addition to inertia.

## Notes and References